# One Side-Graphene Hydrogenation (Graphone): Substrate Effects

Cristiano Francisco Woellner[1], Pedro Alves da Silva Autreto[1,2], Douglas S. Galvao[1]
[1]Instituto de Física "Gleb Wataghin", Universidade Estadual de Campinas, Campinas - SP, 13083-970, Brazil
[2]Universidade Federal do ABC, Santo André-SP, 09210-580, Brazil

## ABSTRACT

Recent studies on graphene hydrogenation processes showed that hydrogenation occurs via island growing domains, however how the substrate can affect the hydrogenation dynamics and/or pattern formation has not been yet properly investigated. In this work we have addressed these issues through fully atomistic reactive molecular dynamics simulations. We investigated the structural and dynamical aspects of the hydrogenation of graphene membranes (one-side hydrogenation, the so called graphone structure) on different substrates (graphene, few-layers graphene, graphite and platinum). Our results also show that the observed hydrogenation rates are very sensitive to the substrate type. For all investigated cases, the largest fraction of hydrogenated carbon atoms was for platinum substrates. Our results also show that a significant number of randomly distributed H clusters are formed during the early stages of the hydrogenation process, regardless of the type of substrate and temperature. These results suggest that, similarly to graphane formation, large perfect graphone-like domains are unlikely to be formed. These findings are especially important since experiments have showed that cluster formation influences the electronic transport properties in hydrogenated graphene.

## INTRODUCTION

Since its experimental realization, graphene [1] became one of the most important subjects in materials science due to its unique and extraordinary electronic, thermal and mechanical properties. However, in its pristine form graphene is a gapless semiconductor, which limits its use in several technological applications. There are many ways to address this gap problem, for example through chemical functionalization (hydrogenation and fluorination, among others). Graphane [2-4] and graphone [5] are the most studied hydrogenated graphene forms. Graphane is the graphene fully hydrogenated form, while graphone has just one hydrogenated side. Graphone presents interesting electronic and mechanical properties, such as non-zero electronic gap and high Young's modulus. These properties can be exploited for different applications in nanoelectronics. Recent studies on graphene hydrogenation processes showed that hydrogenation occurs via island growing domains [3], however how the substrate can affect the hydrogenation dynamics and/or formation patterns has not been yet properly investigated.

In this work we have addressed these issues through fully atomistic reactive molecular dynamics simulations. We investigated the structural and dynamical aspects of the hydrogenation of graphene membranes (graphone-like) on different substrates (graphene, few-layers graphene, graphite and platinum) and at different temperatures.

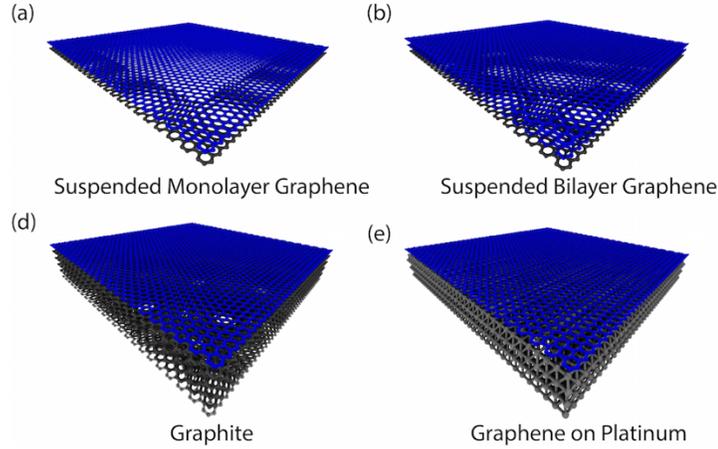

**Figure 1:** Studied systems: (a) Suspended Monolayer Graphene; (b) Suspended Bilayer Graphene; (c) Graphite, and; (d) Graphene on Platinum.

**THEORY**

In order to study the effects of substrate in the one-side graphene functionalization, we have performed fully atomistic molecular dynamics simulations using the reactive force field (ReaxFF) [6], as integrated in the open source code Large-scale atomic/ Molecular Massively Parallel Simulator (LAMMPS) [7]. ReaxFF was developed by van Duin, Goddard III and co-workers and it is used in simulations involving classical molecular dynamics. This force field has the advantage of allowing an accurate description of chemical processes through breaking and formation bonds. Its relative (in comparison to *ab initio* methods) low computational cost makes it applicable to large systems and long simulation times.. Similarly to non-reactive force fields such as MM3 [8], the total energy is composed of several energy terms, such as: bond stretching, bond angle bending, van der Waals and Coulomb interactions, among others, and it can be written in the following form:

$E_{system} = E_{bond} + E_{over} + E_{under} + E_{lp} + E_{val} + E_{pen} + E_{tors} + E_{conj} + E_{vdWaals} + E_{Coulomb}$

A detailed description of all terms can be obtained from Ref. 6. The parameterization of ReaxFF is carried out using Density Functional Theory (DFT) calculations and its accuracy, compared to experimental data, is around 2.9 kcal/mol for unconjugated and conjugated systems [6]. This method has been successfully applied to many investigations of nanomaterials at the atomic scale, including chemical functionalization [3,9].

All molecular dynamics simulations were performed under NVT (canonical) ensemble, using a Nosé-Hoover thermostat [10]. Typical time simulation was around 1 ns, with time steps of 0.1 fs. Since chemical functionalization as hydrogenation is unlikely to occur at low temperatures (even at room temperature), we carried out our simulations from 450 up to 650 K. Due to space limitation, we will present only the results for the case of T=650 K. In the temperature range considered here the general trends are preserved.

In addition to suspended graphene (Figure 1a), we have also considered graphene desposited on three different substrates: graphene (Figure 1b), forming a bilayer system, graphite

(Figure 1c) and platinum (Figure 1d). We mimic the conditions for suspended graphene using a support with a hole in its center. For graphene on graphene, all atoms are free to move and we used again a support with a hole. For all systems, hydrogenation was allowed to occur only at the center of the structure to avoid edges effect, as illustrated in Figure 2.

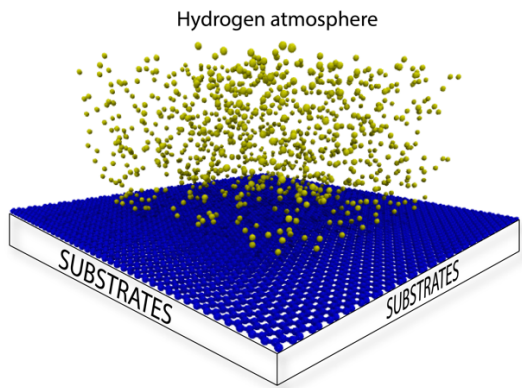

**Figure 2:** Hydrogenation scheme of graphene on substrates. The hydrogen atoms (yellow) are contained in a confined area above graphene membrane to avoid spurious edge effects.

The van der Waals (VdW) interactions between graphene and the substrates can affect both the electronic, as well as, mechanical aspects (deformations) of the membrane. These effects will be crucial to hydrogenation pattern and spatial coverage. One way to obtain static information on these aspects is through potential energy maps. These maps allow us to obtain further insights on preferential hydrogenation sites and to easily evaluate substrate effects. These maps were calculated considering a potential energy variation experienced by a probe hydrogen atom placed at 1.5 Å above the graphene basal plane for each system: suspended and on substrates. Large negative values mean that the regions are energetically favorable for hydrogenation and/or chemical bond formation.

## DISCUSSION

In Figure 3 we present the maps for the different cases. From the potential maps it is possible to identify equivalent sites. For the pristine graphene there are many equivalent sites due to the hexagonal symmetry. However, this equivalence is broken due to substrate and incommensurability effects. These effects are more pronounced to suspended graphene (Figure 3a) and graphene on platinum (Figure 3d). As we can see from Figure 3 this creates more attractive regions for hydrogenation and a greater final hydrogen coverage is expected for these cases. This is corroborated by the results shown in Figure 4 where we presented the hydrogen/carbon atom ratio (effective hydrogenation) as a function of time. As we can see from Figures 3 and 4 the hydrogenation dynamics is quite distinct for the different cases. Graphite and bilayer systems, present lower hydrogenation rates, while hydrogenation coverages reach almost 40% of available carbon atoms for suspended graphene and graphene on platinum.

The hydrogenation processes through the formation of non-correlated island domains observed for graphane [3] is also observed for graphone hydrogenation as can be seen from

Figure 5, in which we presents four representative MD snapshots of the graphene hydrogenation on platinum. The observed cluster formation is substrate independent..

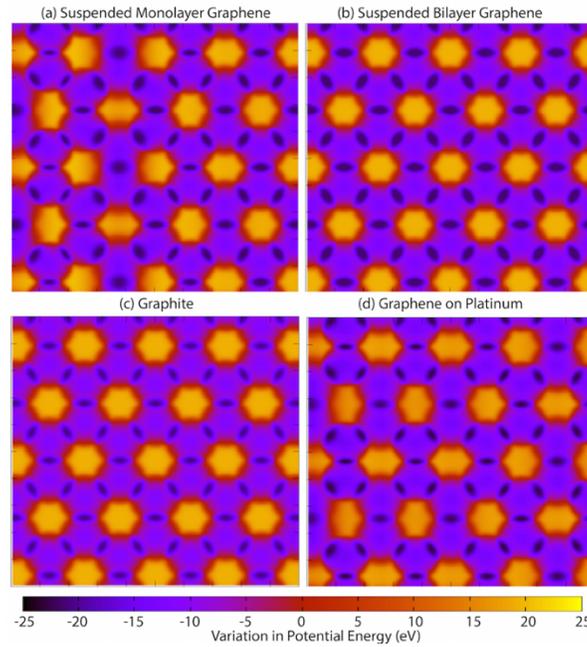

**Figure 3:** Potential energy maps for the different cases investigated here.

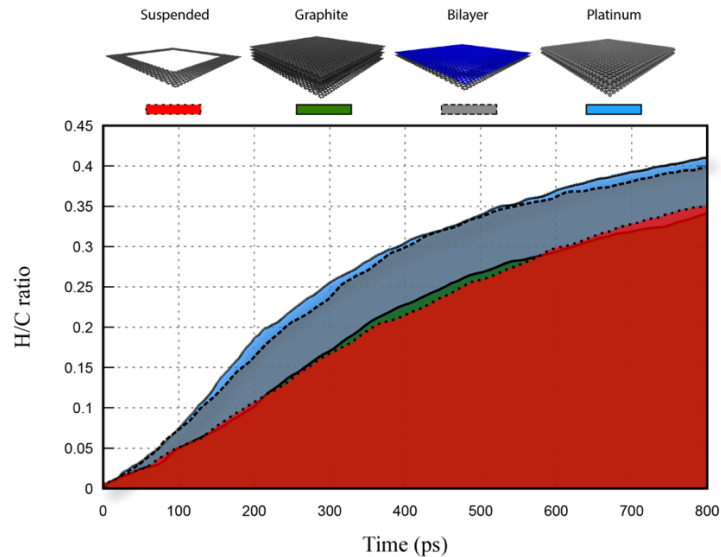

**Figure 4:** Hydrogenation rate as function of simulation time for suspended graphene and graphene on different substrates: graphite, graphene (bilayer system) and platinum. Results for T=650 K.

MD simulations also show that the idealized graphone structures, formed by a complete or almost complete hydrogen coverage, is unlike to be obtained, similarly to the case of graphanes [3]. A more realistic graphone structure would be the co-existence of uncorrelated (randomly distributed) graphone-like islands (Figure 4). It has been argued that graphone could be exploited in some spintronics applications [11]. Our simulations also suggested that the assumed well formed magnetic domains are unlikely to be formed and survive high temperatures. Thus, graphone spintronics applications will be very difficult to be experimentally realized.

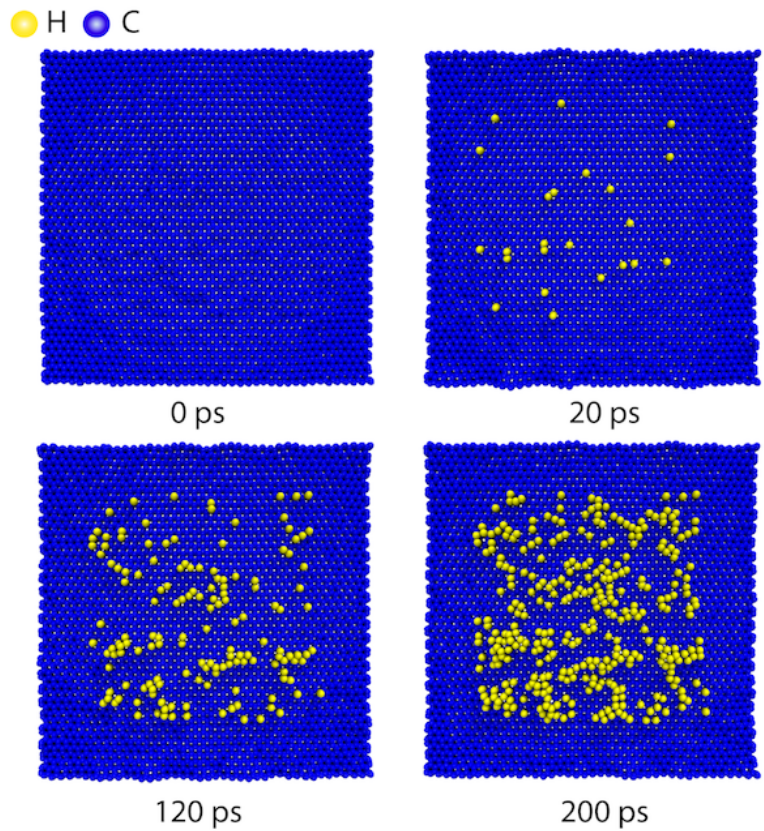

**Figure 5:** Representative MD snapshots of graphene hydrogenation on platinum substrate. The formation of non-correlated island domains were observed in all investigated cases.

**CONCLUSIONS**

We have investigated through fully atomistic reactive molecular dynamics simulations the structural changes and hydrogenation dynamics of one-side graphene (graphone) membranes on different substrates. For comparison purposes graphene freestanding membranes were also considered. Our results show that a significant number of randomly distributed hydrogen clusters are formed during the early stages of the hydrogenation processes, regardless of the type of substrate and temperature. These findings suggest that, similarly to graphane formation [3], large perfect graphone-like domains are unlikely to be formed. Our results also show that the use of graphone in some spintronics (requiring magnetic ordering) applications, as recently proposed [11], will be very difficult to be realized.

ACKNOWLEDGMENTS

The authors acknowledge the São Paulo Research Foundation (FAPESP) Grant No. 2014/24547-1 for financial support. Computational and financial support from the Center for Computational Engineering and Sciences at Unicamp through the FAPESP/CEPID Grant No. 2013/08293-7 is also acknowledged.